   \documentclass[review,12pt]{elsarticle}

\usepackage{amssymb}
\usepackage{booktabs}
\usepackage{amsthm}
\usepackage{amsmath,amsfonts}
\usepackage{times}
\usepackage{lineno,hyperref}

\journal{arXiv}

\begin{document}

\begin{frontmatter}



\title{Closed-Form Formulas for Designing Ultra-Low Phase-Noise Cross-Coupled Dynamically Body-Biased Only-NMOS LCVCOs}


\author[add1]{Naser Khatti Dizabadi \corref{cor1}}\ead{nak5300@utulsa.edu}\cortext[cor1]{Corresponding author}
\author[add1]{Peter LoPresti}\ead{peter-lopresti@utulsa.edu}

\affiliation[add1]{organization={Department of Electrical and Computer Engineering, The University of Tulsa},
            addressline={800 S. Tucker Drive}, 
            city={Tulsa},
            postcode={74104}, 
            state={OK},
            country={USA}}

\begin{abstract}
This paper presents a system-level analytical framework for modeling and minimizing phase noise in body-biased cross-coupled LC-tank voltage-controlled oscillators (LC-VCOs). Building upon Impulse Sensitivity Function (ISF) theory, the impulse sensitivity and noise modulation mechanisms associated with both flicker and thermal noise sources are systematically characterized. By modeling the oscillator as a nonlinear dynamical system and incorporating transistor operation across multiple regions, analytical expressions for device-level noise power spectral densities (PSDs) are derived as functions of transconductance parameters under symmetric body excitation. Using these results, effective ISF representations corresponding to dominant noise sources are formulated, enabling a unified description of noise-to-phase conversion dynamics. The phase noise minimization problem is then cast as an optimization over system parameters, where both DC and RMS components of the effective ISF are analytically evaluated and minimized. This leads to the derivation of three closed-form expressions that explicitly capture the interaction between circuit parameters and the applied body-bias signals. The proposed framework provides insight into parameter sensitivity and design trade-offs in nonlinear oscillator systems and offers generalizable analytical tools for guiding the design of ultra-low phase noise LC-VCOs, as well as for exploring new oscillator architectures.

\end{abstract}




\end{frontmatter}


\section{Introduction}
\label{intr}
Wireless transceivers rely on oscillators to generate high-quality, high-frequency carrier signals, minimizing bandwidth wastage across diverse frequency spectra \cite{VCO-A1-0}. Achieving such high-quality carrier signals requires local oscillators to generate ultra-low phase noise sinusoidal waveforms, which facilitate ultra-narrow frequency response.
In many transceiver designs, voltage-controlled oscillators (VCOs) serve as the core component of local oscillators \cite{VCO-A1-00}. Therefore, it is imperative to understand the phase noise characteristics of VCOs to identify circuit parameters that contribute to overall phase noise at the output oscillating nodes.
Among the various types of oscillators, the LC Tank VCO (LC-VCO) has superior phase noise performance, making it a preferred choice for low-noise applications. To implement low phase noise LC-VCOs, a variety of methods and circuit topologies are employed, such as current reuse, current shaping, tank-tuned LCs, body-biased LCs, class-Cs, filters, and inverse class-Fs \cite{VCO-A1-11, VCO-A1-12, VCO-A1-13, VCO-A1-9, VCO-A1-10, VCO-A1-16, VCO-A1-17, VCO-A1-14, VCO-A1-15}. 
Among these techniques, the body biasing technique utilizes the body pins of cross-coupled transistors as back gates to have more control over the drain currents of the devices. This can be achieved by nonlinearly regulating core transistor threshold voltages. However, efficient control of body signals in the body biasing technique necessitates comprehensive mathematical analysis to discern the impact of circuit parameters on phase noise performance. In particular, flicker and thermal noise generated by cross-coupled transistors emerge as key noise sources in cross-coupled LC-VCOs.\\ 
In this paper, a comprehensive mathematical analysis aimed at minimizing phase noise is conducted on only-NMOS dynamically body-biased LC-VCOs. Through this analysis, three closed-form formulas are derived for both the DC and AC components of the signals applied to the body pins of the cross-coupled transistors. 
When the minimization of phase noise due to both flicker and thermal noise from active devices is targeted, these formulas can be used to determine how the DC and AC components of the body signals should be tuned. It is also possible to introduce different circuit topologies that satisfy these formulas.
As a robust theory for analyzing phase noise, Hajimiri's linear Time-Variant (LTV) phase noise model provides insights into the contribution of thermal and flicker noise sources to phase noise at the output nodes of an oscillator. As a result of the Impulse Sensitivity Function (ISF) within this theory, it is possible to reduce phase noise by minimizing the DC and RMS values of the ISF functions associated with thermal and flicker noise sources \cite {VCO-A1-7, VCO-A1-8}. Accordingly, the phase noise analyses in this study are grounded in Hajimiri’s ISF theory.\\ 
The remaining portions of this paper are organized as follows: Section 2 employs a three-terminal model for NMOS transistors to calculate relative transconductance, which permits the formulation of Power Spectral Density (PSD) of flicker and thermal noise sources in Section 3. Section 4 looks at the different work regions of the cross-coupled transistors and examines how variations in the gate and body voltages can affect oscillator behavior. Section 5 discusses phase noise in a conventional cross-coupled only-NMOS LC-VCO, followed by sections 6 and 7 examining the DC and RMS values of the effective ISF functions derived from ISF theory. The intent of Section 8 is to provide optimal angles for the borders of the working regions of NMOS transistors so that overall phase noise is minimized. Section 9 summarizes three closed-form formulas which parametrically describe the AC amplitude and DC level of the symmetric body signals and how they interact with other parameters to minimize phase noise. As discussed in Section 10, a feedback-based Simulink model is presented to demonstrate phase noise reduction when body terminals are dynamically biased. Finally, Section 11 provides conclusions.

\section{Transconductance in a Three-Terminal NMOS Transistor}
\label{ISF-PN}
\begin{figure}[]
\centering
\noindent
\includegraphics[width=3.5in]{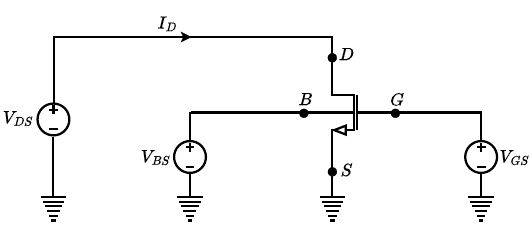}
\caption{A three-terminal NMOS transistor}
\label{fig-02}
\end{figure}

In a typical NMOS transistor with a grounded source pin, as illustrated in figure \ref{fig-02}, the drain-source current and relative transconductance are a function of three variables: the gate-source, drain-source, and bulk-source voltages.

\begin{figure}[]
\centering
\noindent
\includegraphics[width=3.5in]{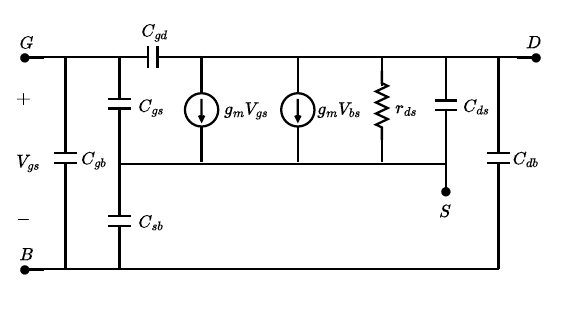}
\caption{AC equivalent circuit of a three-terminal NMOS transistor}
\label{fig-03}
\end{figure}

Although the parasitic capacitances formed between the different terminals of a CMOS transistor can be ignored at low frequencies, they can cause major problems at high frequencies by creating unwanted feedback and undesirable effects. Figure \ref{fig-03} shows the AC equivalent circuit for the NMOS transistor represented in figure \ref{fig-02}. 

According to this model, the drain current of an NMOS transistor can generally be described by the following equation:

\begin{equation}
\label{01}
I_{D} = F(V_{GS}, V_{DS}, V_{BS})
\end{equation}

where \(I_{D}\) is the drain-source current, \(V_{GS}\) is the gate-source voltage, \(V_{DS}\) is the drain-source voltage, and \(V_{BS}\) is the bulk-source voltage. In order to express the transconductance (\(g_{m}\)) of a three terminal NMOS transistor, the following expression of three partial derivatives is used. 

\begin{equation}
\label{02}
g_{m} = \frac{\partial{I_{D}}}{\partial{V_{GS}}} + \frac{\partial{I_{D}}}{\partial{V_{BS}}} + \frac{\partial{I_{D}}}{\partial{V_{DS}}}
\end{equation}

As reported in \cite{VCO-A1-19}, the relationship between threshold voltage (\(V_{Th}\)) and bulk-source voltage (\(V_{BS}\)) can be expressed as follows:

\begin{equation}
\label{03 - A}
V_{Th} = V_{Th0} + \gamma (\sqrt{\Phi_{s} - V_{BS}} - \sqrt{\Phi_{s}})
\end{equation}

where \(V_{Th0}\) is the threshold voltage when \(V_{BS}=0\), \(\gamma\) is the body-effect coefficient, \(\Phi_{s}\) is the potential at the surface of the semiconductor relative to the bulk or substrate, and the \(V_{BS}\) is the bulk voltage with respect to the source. To avoid unnecessary calculations originating from the complex nature of Equation (\ref{03 - A}), the Maclaurin series expansion can be used as a logical approximation of the threshold voltage around \(V_{BS}=0\). Therefore, using Maclaurin series expansion, the Equation (\ref{03 - A}) can be rewritten as described below: 

\begin{equation}
\label{03 - B}
V_{Th} = V_{Th0} + \sum_{k=1}^{\infty} \frac{(-1)^k}{k!} \frac{\gamma}{\sqrt {\Phi_{s}^{(2k-1)}}}V_{BS}^{k}
\end{equation}

Due to the fact that the bulk-source voltage range in integrated circuits is limited, Equation (\ref{03 - B}) can be approximated based on only the first derivative (k=1) of the Maclaurin series without introducing significant errors. Therefore, by updating the Equation (4), the threshold voltage of an NMOS transistor can be approximated by the following formula:

\begin{equation}
\label{03}
V_{Th} = V_{Th0} - n V_{BS}
\end{equation}

where the factor \(n\) is

\begin{equation}
\label{04}
n = \frac{\gamma}{\sqrt {\Phi_{s}}}.
\end{equation}

Nonetheless, it should be noted that this approximation will not be valid when the bulk-source voltage becomes larger than \(\Phi_{s}\).

\subsection{Transconductance in the Saturation Region}
\label{gm-S}

According to Equation (\ref{02}), an equation expressing the behavior of the drain current of a three-terminal NMOS transistor in saturation mode is needed to calculate its transconductance. Thus, assuming the channel length modulation effect is approximately ignored in a NMOS transistor, the following drain current equation \cite{VCO-A1-19} is used.

\begin{equation}
\label{06}
I_{D-S} = \frac{1}{2}\mu_{n}C_{ox}(\frac{w}{l})(V_{GS}-V_{Th})^2 
\end{equation}

where \(I_{D-S}\) is the drain current in saturation mode, \(C_{ox}\) is the oxide capacitance, \(\mu_{n}\) is the mobility of electrons, \(\textit w\) is the width of the transistor, and \(\textit l\) is the length of the transistor. By substituting the \(V_{Th}\) from equation (\ref{03}) into equation (\ref{06}), the drain current in saturation mode can be updated as follows:

\begin{equation}
\label{07}
I_{D-S} = \frac{1}{2}\mu_{n}C_{ox}(\frac{w}{l})(V_{GS}-V_{Th0} + nV_{BS})^2 
\end{equation}

Assuming that \(\textit r_{ds}=\infty\), it is evident from equation (\ref{07}) that the drain current in this region is independent of the drain-source voltage, so, in the saturation region, equation (\ref{02}) can be revised as follows:

\begin{equation}
\label{08}
g_{m-S} = \frac{\partial I_{D}}{\partial V_{GS}} + \frac{\partial I_{D}}{\partial V_{BS}}
\end{equation}

where \(g_{m-s}\) is the transistor’s transconductance in saturation mode. Now, by applying equation (\ref{07}) to equation (\ref{08}), the transconductance of a three-terminal NMOS in saturation mode is given below:

\begin{equation}
\label{09}
g_{m-S} = \mu_{n}C_{ox}(\frac{w}{l})(n+1)[V_{GS} + nV_{BS} - V_{Th0}]
\end{equation}

\subsection{Transconductance in the Triode Region}
\label{gm-T}
In a manner similar to saturation mode, equation (\ref{02}) can also be utilized to calculate the transconductance of a three-terminal NMOS transistor in triode mode if the following equation describing the drain current behaviour in this work region is used \cite{VCO-A1-19}.

\begin{equation}
\label{010}
I_{D-T} = \mu_{n}C_{ox}(\frac{w}{l})[(V_{GS} - V_{Th}) - \frac{V_{DS}}{2}]V_{DS}
\end{equation}

where \(I_{D-T}\) is the drain current in triode mode. By substituting \(V_{Th}\) from equation (\ref{03}) into equation (\ref{010}), the drain current in triode mode can be updated as follows:

\begin{equation}
\label{011}
I_{D-T} = \mu_{n}C_{ox}(\frac{w}{l})[(V_{GS} - V_{Th0} + nV_{BS}) - \frac{V_{DS}}{2}]V_{DS}
\end{equation}

Equation (\ref{011}) shows that the drain current in the triode region is not only affected by the gate-source voltage but also by the bulk-source and drain-source voltages. In this case, the relative transconductance of the transistor in triode mode can be derived by applying equation (\ref{011}) to equation (\ref{02}):

\begin{equation}
\label{012}
g_{m-T} = \mu_{n}C_{ox}(\frac{w}{l})[V_{GS} + n(V_{DS} + V_{BS}) - V_{Th0}] 
\end{equation}

where \(g_{m-s}\) is the transistor’s transconductance in triode mode.

\section{PSD of Noise Sources in a Three-Terminal NMOS Transistor}
\label{PSD}
A CMOS transistor can be thought of as a combination of noise sources. Flicker and thermal noise are two of the most prominent noise sources in a typical CMOS transistor \cite{VCO-A1-20}. According to \cite{VCO-A1-21, VCO-A1-22}, the following equations illustrate the relationship between the transistor's transconductance and its intrinsic flicker and thermal noise PSDs. 

\begin{equation}
\label{013}
\frac{\bar{i}_{n-f}^2}{\Delta f} = \frac{K}{C_{ox}wl}\frac{1}{f} g_{m}^2
\end{equation}

\begin{equation}
\label{014}
\frac{\bar{i}_{n-th}^2}{\Delta f} = 4KT \gamma g_{m}
\end{equation}

where \(\bar{i}_{n-f}^2/\Delta f\) is the PSD of the flicker noise, \(\bar{i}_{n-th}^2/\Delta f\) is the PSD of the thermal noise, K is the Boltzmann's constant, f is the frequency, T is the temperature, \(\gamma\) is the channel thermal noise coefficient, and \(g_{m}\) is the transistor’s transconductance. It is clear from the above equations that to minimize or manipulate the PSDs of flicker and thermal noise sources in a three-terminal NMOS transistor, the transconductance of the transistor plays a substantial role. Since an NMOS transistor can have different transconductance in its different work regions, the PSD of these two noise sources must be calculated separately in the saturation and triode regions, while, in the cut-off region, the transistor is assumed not to generate any noise and its PSD is essentially zero.

\subsection{PSD of Flicker and Thermal Noise Sources in the Saturation Region}
\label{PSD-s}
In the saturation region of a three-terminal NMOS transistor, substituting equation (\ref{09}) into equations (\ref{013}) and (\ref{014}) yields the following expressions for the power spectral density (PSD) of flicker and thermal noise sources:

\begin{equation}
\label{015}
\frac{\bar{i}_{n-f_{S}}^2}{\Delta f} = \frac{K\mu_{n}^2C_{ox}w(n+1)^2}{l^3f}[V_{GS} + nV_{BS} - V_{th0}]^2
\end{equation}

\begin{equation}
\label{016}
\frac{\bar{i}_{n-th_{S}}^2}{\Delta f} = \frac{KT\gamma\mu_{n}C_{ox}w(n+1)}{l}[V_{GS} + nV_{BS} - V_{th0}]
\end{equation}

where \(\bar{i}_{n-f_{S}}^2/\Delta f\) and \(\bar{i}_{n-th_{S}}^2/\Delta f\) are the PSD of flicker and thermal noise in saturation mode, respectively.

\subsection{PSD of Flicker and Thermal Noise Sources in the Triode Region}
\label{PSD-T}

When operating in the triode region, the PSD expressions for flicker and thermal noise sources in a three-terminal NMOS transistor can be obtained by substituting equation (\ref{012}) into equations (\ref{013}) and (\ref{014}):

\begin{equation}
\label{017}
\frac{\bar{i}_{n-f_{T}}^2}{\Delta f} = \frac{K\mu_{n}^2C_{ox}w}{l^3f}[V_{GS} + n(V_{DS} + V_{BS}) - V_{th0}]^2
\end{equation}

\begin{equation}
\label{018}
\frac{\bar{i}_{n-th_{T}}^2}{\Delta f} = \frac{4KT\gamma\mu_{n}C_{ox}w}{l}[V_{GS} + n(V_{DS} + V_{BS}) - V_{th0}]
\end{equation}

where \(\bar{i}_{n-f_{T}}^2/\Delta f\) and \(\bar{i}_{n-th_{T}}^2/\Delta f\)  are the PSD of the flicker and thermal noise in triode mode, respectively.

\section{Transistor Work Regions in an Only-NMOS Cross-Coupled LC-VCO}
\label{Regions}
As shown by equations (\ref{015}), (\ref{016}), (\ref{017}), and (\ref{018}), changing the gate-source, drain-source, and bulk-source voltages of a three-terminal NMOS transistor results in a change in the PSD of flicker and thermal noise. These PSDs, however, vary depending on the regions where transistors operate. The work regions of a transistor must therefore be analyzed during a given oscillation period of length T. By calculating the borders between these regions, a transistor's work regions can be distinguished. It is at these borders that a transistor switches from one mode to another. Considering that the work regions of two transistors in an only-NMOS cross-coupled LC-VCO may have a correlation, it is beneficial to analyze their work regions while the oscillator is in steady state. 

\begin{figure}[]
\centering
\noindent
\includegraphics[width=3.5in]{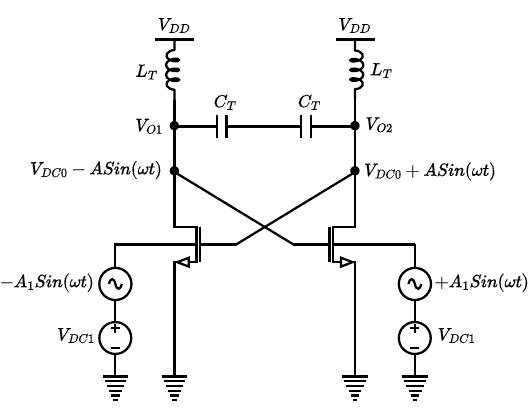}
\caption{A conventional body-biased only-NMOS cross-coupled LC-VCO}
\label{fig-04}
\end{figure}

In Figure \ref{fig-04}, a cross-coupled only-NMOS LC-VCO is shown. Transistors in this figure generate negative resistance to cancel out the positive loss of the LC tank, resulting in sinusoidal oscillations at the output. It is necessary to first characterize all the circuit parameters in order to mathematically extract the angles at which the work regions are changed. Inductor and capacitor values are factors that determine the frequency of oscillation. The body terminals of the transistors are dynamically biased using a combination of AC and DC signals. The AC components consist of two symmetrical sinusoidal waveforms with amplitude A1, operating at the same frequency as the LC tank but 180 degrees out of phase with each other. The DC component, denoted by \(V_{DC1}\), sets the baseline bias level for the body terminals. Assuming the LC-VCO is functioning in its steady state, the output oscillating voltages and body signals can be formulated as follows:

\begin{equation}
\label{019}
\left\{
\begin{array}{ll}
V_{DS1} = V_{O1} = V_{DC0} - A sin(\omega t) \\
V_{GS1} = V_{O2} = V_{DC0} + A sin(\omega t) \\
V_{BS1} = V_{DC1} - A_{1} sin(\omega t)
\end{array}
\right.
\end{equation}

\begin{equation}
\label{020}
\left\{
\begin{array}{ll}
V_{DS2} = V_{O2} = V_{DC0} + A sin(\omega t) \\
V_{GS2} = V_{O1} = V_{DC0} - A sin(\omega t) \\
V_{BS2} = V_{DC1} + A_{1} sin(\omega t)
\end{array}
\right.
\end{equation}

\subsection{Saturation Region}
\label{S}
In figure \ref{fig-04}, the cross-coupled feedback mechanism in the oscillator is intended to provide adequate negative resistance to counterbalance the parallel positive resistance observed in the LC tank. As a result of this synergy, oscillation can be initiated within the oscillator circuit and reach its steady state after passing a start-up condition. During the steady state, M1 and M2 will keep striving to provide the required currents for maintaining the average negative resistance while constantly changing their work regions over a period of length T. Simultaneously, the transistors' body pins are supplied by two external symmetric voltages, \(V_{B1}\) and \(V_{B2}\), to dynamically manipulate their threshold voltages. When evaluating a typical NMOS transistor in its saturation region, the transistor must first be on, which means its gate-source voltage must be greater than its threshold voltage. The next step for the transistor to remain in saturation mode is that the drain-source voltage must be equal to or greater than the overdrive voltage (\(V_{GS}-V_{Th})\). In general, taking equation (\ref{03}) into account, body-biased NMOS transistors must meet the following conditions in order to work in saturation mode:
 
\begin{equation}
\label{021}
V_{GS} \geq V_{Th0} - nV_{B}
\end{equation}

\begin{equation}
\label{022}
V_{DS} \geq V_{GS} - (V_{Th0} - nV_{B})
\end{equation}

For the M1 transistor, since the drain, gate, and body pins are controlled by \(V_{O1}\), \(V_{O2}\), and \(V_{B1}\), the saturation criteria can be updated as follows:

\begin{equation}
\label{023}
V_{O2} \geq V_{Th0} - nV_{B1}
\end{equation}

\begin{equation}
\label{024}
V_{O1} + (V_{Th0} - nV_{B1}) \geq V_{O2}
\end{equation}

Now, by applying equation (\ref{019}) to equations (\ref{023}) and (\ref{024}) and after mathematical simplifications, the following criteria must be satisfied for M1 to work in its saturation region.

\begin{equation}
\label{025}
sin(\omega t) \geq \frac{V_{Th0} - nV_{DC1} - V_{DC0}}{A - nA_{1}}
\end{equation}

\begin{equation}
\label{026}
sin(\omega t) \leq \frac{V_{Th0} - nV_{DC1}}{2A - nA_{1}}
\end{equation}

The following border angles can be determined from equations (\ref{025}) and (\ref{026}):

\begin{equation}
\label{027}
\phi_{1} = sin^{-1}(\frac{V_{Th0} - nV_{DC1} - V_{DC0}}{A - nA_{1}})
\end{equation}

\begin{equation}
\label{028}
\phi_{2} = sin^{-1}(\frac{V_{Th0} - nV_{DC1}}{2A - nA_{1}})
\end{equation}

where \(\phi_{1}\) is the angle at which M1 switches its mode from OFF to ON (or ON to OFF) and \(\phi_{2}\) is the angle at which M1 changes its working mode from saturation to triode and vice versa. By rewriting equations (\ref{025}) and (\ref{026}) based on \(\phi_{1}\) and \(\phi_{2}\), the following expression is obtained for the M1 transistor in saturation mode.

\begin{equation}
\label{029}
sin(\phi_{1}) \leq sin(\omega t) \leq sin(\phi_{2})
\end{equation}

After solving equation (\ref{029}), M1 operates in saturation mode over the following ranges for a period of \(2\pi\).

\begin{equation}
\label{030}
\left\{
\begin{array}{ll}
0 \leq \omega t \leq \phi_{2}\\
\pi - \phi_{2} \leq \omega t \leq \pi + \phi_{1}\\
2\pi - \phi_{1} \leq \omega t \leq 2\pi
\end{array}
\right.
\end{equation}

\subsection{Triode Region}
\label{T}

As in the saturation mode, the device must first be turned on when operating in the triode region. Triode operation requires the drain-source voltage \((V_{DS})\) to be less than the overdrive voltage \((V_{GS} - V_{Th})\). Based on this condition, equations (\ref{025}) and (\ref{026}) are modified to take the following form:

\begin{equation}
\label{031}
sin(\omega t) \geq \frac{V_{Th0} - nV_{DC1} - V_{DC0}}{A - nA_{1}}
\end{equation}

\begin{equation}
\label{032}
sin(\omega t) > \frac{V_{Th0} - nV_{DC1}}{2A - nA_{1}}
\end{equation}

For the M1 transistor in triode mode, the following expression can be obtained by rewriting equations (\ref{031}) and (\ref{032}) based on the border angles defined in (\ref{027}) and (\ref{028}).

\begin{equation}
\label{033}
\left\{
\begin{array}{ll}
sin(\omega t) \geq sin(\phi_{1})\\
sin(\omega t) > sin(\phi_{2})
\end{array}
\right.
\end{equation}

Solving equation (\ref{033}) yields the following range, which defines the interval over which M1 operates in the triode region throughout the entire \(2\pi\) cycle.

\begin{equation}
\label{034}
\phi_{2} < \omega t < \pi - \phi_{2}
\end{equation}

\subsection{Cutt Off Region}
\label{C}
If M1 operates in the cut-off region when its gate-source voltage falls below the threshold voltage, equation (\ref{025}) must be violated. Therefore, for this operating region, the equation takes the following form:

\begin{equation}
\label{035}
sin(\omega t) < \frac{V_{Th0} - nV_{DC1} - V_{DC0}}{A - nA_{1}}
\end{equation}

Taking equation (\ref{027}) into account, equation (\ref{035}) is revised as follows:

\begin{equation}
\label{036}
sin(\omega t) < sin(\phi_{1})
\end{equation}

From equation (\ref{036}), the following range can be derived, indicating where M1 operates in the cut-off mode.

\begin{equation}
\label{037}
\pi + \phi_{1} < \omega t < 2\pi - \phi_{1}
\end{equation}

\begin{figure}[]
\centering
\noindent
\includegraphics[width=3.5in]{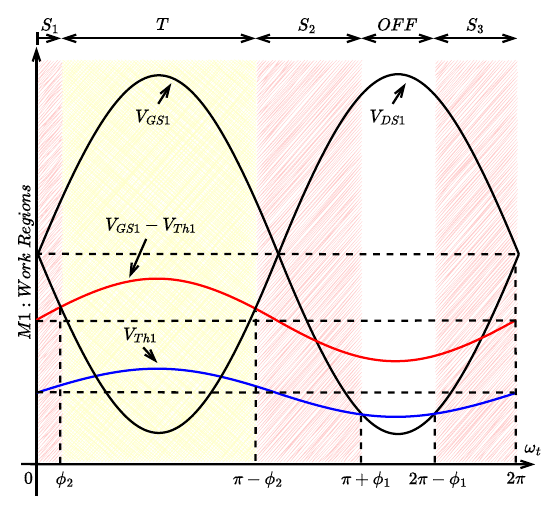}
\caption{Different Work Regions of \(M_{1}\). The red areas represent the saturation regions, denoted as \textbf{S}, the yellow area indicates the triode region, labeled as \textbf{T}, and the white area corresponds to the cut-off region, marked as \textbf{OFF}.}
\label{fig-05}
\end{figure}

\begin{figure}[]
\centering
\noindent
\includegraphics[width=3.5in]{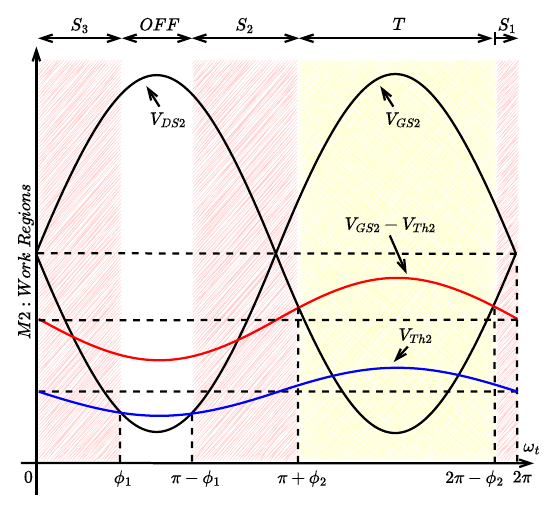}
\caption{Different Work Regions of \(M_{2}\). The red areas represent the saturation regions, denoted as \textbf{S}, the yellow area indicates the triode region, labeled as \textbf{T}, and the white area corresponds to the cut-off region, marked as \textbf{OFF}.}
\label{fig-05-2}
\end{figure}

Similarly, Appendix (A) discusses the work regions of M2. Due to the symmetrical design of the oscillator shown in Figure~\ref{fig-04}, the saturation, triode, and cut-off regions of M2 are phase-shifted by \(180^\circ\) relative to those of M1. Figures~\ref{fig-05} and \ref{fig-05-2} illustrate the operating regions of both transistors while the oscillator functions in steady state over a complete \(2\pi\) cycle. In Figure~\ref{fig-05}, the black waveforms represent the gate-source and drain-source voltages of M1. According to equation (\ref{021}), the blue waveform indicates the threshold voltage for M1, while, based on equation (\ref{022}), the red waveform represents its overdrive voltage (\(V_{GS1} - V_{Th1}\)). The blue waveform, derived from equation~(\ref{021}), indicates the threshold voltage, whereas the red waveform, based on equation~(\ref{022}), represents the overdrive voltage \((V_{GS1} - V_{Th1})\). During one full cycle, M1 operates in the saturation region within three distinct intervals labeled S1, S2, and S3. At the angle of \(\phi_{2}\), the mode changes from saturation to triode, then it switches back to saturation at \(\pi - \phi_{2}\). At \(\pi + \phi_1\), if \(V_{GS}\) falls below the dynamic threshold (blue curve), the transistor turns off and remains in the cut-off region until \(2\pi - \phi_1\), when it enters the third saturation region due to \(V_{DS}\) exceeding the overdrive voltage. Figure~\ref{fig-05-2} shows the corresponding waveforms for M2. The black curves indicate \(V_{DS2}\) and \(V_{GS2}\), while the blue and red curves represent the threshold and overdrive voltages, respectively, derived using the equations (\ref{021}) and (\ref{022}). M2 also operates in three saturation intervals. The first occurs from 0 to \(\phi_1\), the second from \(\pi - \phi_1\) to \(\pi + \phi_2\), and the third from \(2\pi - \phi_2\) to \(2\pi\). In the interval \([\phi_1, \pi - \phi_1]\), M2 is in the cut-off region, and between \([\pi + \phi_2, 2\pi - \phi_2]\), it operates in triode mode.

\section{Phase Noise Analysis Based on ISF Theory}
\label{ISF-PN}
This section examines the phase noise of the body-biased only-NMOS LC-VCO shown in figure \ref{fig-04}. Due to the inherent nonlinear behavior of each transistor in different work regions, the phase noise induced by flicker and thermal noise sources of each transistor can be modulated during the process of generating negative resistance, thereby complicating the analysis of phase noise at the output nodes in this schematic. For a comprehensive analysis of phase noise in cross-coupled LC-VCOs, a precise theory is required to accommodate thermal and flicker noise sources, given the constantly changing work regions of the cross-coupled transistors. The ISF theory \cite{VCO-A1-7} is one of the most accurate theories for analyzing phase noise in LC oscillators. This theory describes the interaction between oscillator output signals and thermal and flicker noise sources as a cyclo-stationary linear time-varying system. The first step to extend this theory to the circuit drawn in figure \ref{fig-04} is calculating the ISF based on the output signals. As reported by \cite{VCO-A1-7}, the following formula can be used to calculate the ISF:

\begin{equation}
\label{038}
\Gamma(\omega t) = \frac{V'_{O}}{{V'_{O}}^2 + {V''_{O}}^2}
\end{equation}

where \(\Gamma(\omega t)\) describes the ISF and \(V_{O}\) represents one of the single-ended sinusoidal output waveforms. The first and second derivatives in Equation (\ref{038}) have been used in such a way as to make a 90-degree phase difference between the \(\Gamma(\omega t)\) and the output signals, showing that when a transistor injects noise into the LC tank, the zero crossing points of the output oscillating signals are the most sensitive points over a period of \(2\pi\). According to \cite{VCO-A1-23}, it is also possible to treat cyclostationary noise as a stationary noise source, and that equation (\ref{038}) can be updated to reflect this:

\begin{equation}
\label{039}
\Gamma_{eff}(\omega t) = \Gamma(\omega t) \times \alpha(\omega t)
\end{equation}

where \(\Gamma_{eff}(\omega t)\) is the effective ISF and \(\alpha(\omega t)\) is a mathematical description of noise amplitude modulation whose magnitude is normalized to a maximum value of 1. This dimensionless function is also known as the Noise Modulating Function (NMF) and illustrates the time when each transistor is ON and generates noise. By applying \(V_{O1}\) and \(V_{O2}\) formulated through equations (\ref{019}) and (\ref{020}) to equation (\ref{038}), we can derive the following ISF functions for each single-ended output node:

\begin{equation}
\label{040}
\Gamma_{1}(\omega t) = -cos(\omega t)
\end{equation}

\begin{equation}
\label{041}
\Gamma_{2}(\omega t) = +cos(\omega t)
\end{equation}

where \(\Gamma_{1}(\omega t)\) is the ISF of \(V_{O1}\) and \(\Gamma_{2}(\omega t)\) is the ISF of \(V_{O2}\).
As reported in \cite{VCO-A1-23}, \(\alpha(\omega t)\) can be determined from the transistor noise characteristics, the PSDs of flicker and thermal noise of a transistor. According to Equations (\ref{013}) and (\ref{014}), CMOS transistor flicker and thermal PSDs are influenced by their transconductance, which can also be affected by the transistor's bulk-source, drain-source, and gate-source signals in different work regions. Therefore, finding appropriate values for the amplitudes and DC offsets of both output oscillating signals and body signals will enable us to optimize the alpha function of each noise source in a transistor to reduce overall phase noise. To accomplish this, the following two sections independently analyze the phase noise originating from both flicker and thermal noise of each NMOS transistor in the LC-VCO shown in figure \ref{fig-04} by calculating parametric alpha functions.

\section{Phase Noise from Flicker Noise}
\label{ISF-f}
The ISF theory states that phase noise from flicker noise can be expressed by the following formula:

\begin{equation}
\label{042}
L(\Delta\omega) \approx 10\log_{10}(\frac{{C_{0}}^2}{8q_{max}^2}\frac{\bar{i}_{n-f}^2}{\Delta f}\frac{\omega_{1/f}}{\Delta {\omega}^3})
\end{equation}

where \(C_{0}\) is the DC value of \(\Gamma_{eff}(\omega t)\), \(q_{max}\) is the maximum charge across the LC tank, \({\bar{i}_{n-f}^2}/{\Delta f}\) is the power spectral density of flicker noise, \(\omega_{1/f}\) is flicker noise corner frequency of a MOS device, and \(\Delta\omega\) is the frequency offset for one flicker noise source. 
According to the above expression, one method of reducing phase noise caused by flicker noise in a MOS device is by minimizing \(C_{0}\). In figure \ref{fig-04}, the first step to calculate \(C_{0}\) for the flicker noise of NMOS transistors is to determine their effective ISF. Assuming that the LC-VCO is designed symmetrically, the results will be similar for M1 and M2. Thus, to avoid redundancy, \(C_{0}\) will be calculated only for M1.

\subsection{Calculating \(\Gamma_{eff}(\omega t)\) due to Flicker Noise of M1 }
\label{effISF-f}
Equation (\ref{039}) implies that before calculating \(\Gamma_{eff}(\omega t)\) for the M1 transistor, in addition to its ISF, \(\Gamma_{1}(\omega t)\), the corresponding \(\alpha(\omega t)\) must also be determined. As M1 constantly changes its working mode throughout a period of oscillation in steady state, \(\alpha(\omega t)\) must be found in three different regions. In the cut-off region, \(g_{m}\) is nearly zero when compared to other regions, so the \(\alpha(\omega t)\) within this region is set to 0. To calculate the \(\alpha(\omega t)\) for the other regions, the following formula reported in \cite{VCO-A1-24} is used:

\begin{equation}
\label{043}
\alpha_{f_{1}}(\omega t) = \frac{\frac{\bar{i}_{n-f}^2}{\Delta f}}{max(\frac{\bar{i}_{n-f}^2}{\Delta f})} 
\end{equation}

where \(\alpha_{f_{1}}(\omega t)\) is the M1 noise modulating function due to flicker noise, and \({\bar{i}_{n-f}^2}/{\Delta f}\) is the flicker noise power spectral density of the device. By substituting Equation (\ref{015}) into Equation (\ref{043}), the following NMF is derived for M1 in the saturation region:

\begin{equation}
\label{044}
\alpha_{f_{1-S}}(\omega t) = \frac{[V_{GS1} + nV_{BS1} - V_{Th0}]^2}{max([V_{GS1} + nV_{BS1} - V_{Th0}]^2)} 
\end{equation}

where \(\alpha_{f_{1-S}}(\omega t)\) represents the noise modulating function caused by the flicker noise of M1 in saturation mode. By applying the assumptions expressed in Equation (\ref{019}) to Equation (\ref{044}), and after mathematical simplifications, the normalized NMF for M1 due to its internal flicker noise in saturation mode can be updated as follows:

\begin{equation}
\label{045}
\alpha_{f_{1-S}}(\omega t) = sin(\frac{\phi_{1} - \omega t}{2})cos(\frac{\phi_{1} + \omega t}{2})
\end{equation}

With Equation (\ref{045}) and (\ref{040}) substituted into (\ref{039}), \(\Gamma_{eff}(\omega t)\) for M1 in saturation mode is as follows:

\begin{equation}
\label{046}
\Gamma_{eff_{f-1-S}}(\omega t) = cos(\omega t) sin(\frac{\phi_{1} - \omega t}{2})cos(\frac{\phi_{1} + \omega t}{2})
\end{equation}

In the triode mode of M1, Equation (\ref{017}) can be substituted into Equation (\ref{043}), resulting in the following NMF:

\begin{equation}
\label{047}
\alpha_{f_{1-T}}(\omega t) = \frac{[V_{GS1} + n(V_{DS1} + V_{BS1}) - V_{Th0}]^2}{max([V_{GS1} + n(V_{DS1} + V_{BS1}) - V_{Th0}]^2)} 
\end{equation}


Assuming that, in an NMOS-only cross-coupled LC-VCO, \(V_{DC0} \approx A \approx V_{DD}\), and applying the assumptions outlined in Equation~(\ref{019}), the normalized NMF for M1 operating in triode mode can be expressed in the following trigonometrically simplified form:

\begin{equation}
\label{048}
\alpha_{f_{1-T}}(\omega t) = sin(\frac{\phi_{X} - \omega t}{2})cos(\frac{\phi_{X} + \omega t}{2})
\end{equation}

where \(\alpha_{f_{1-T}}(\omega t)\) denotes the noise modulation function due to the flicker noise of M1 in triode mode, and \(\phi_{X}\) is defined as

\begin{equation}
\label{049}
\phi_{X} = sin^{-1}(\frac{V_{Th0} - nV_{DC1} - (n+1)V_{DC0}}{(1-n)A - nA_{1}})
\end{equation}

When Equation (\ref{048}) and (\ref{040}) are substituted into Equation (\ref{039}), the \(\Gamma_{eff}(\omega t)\) of M1 in triode mode is obtained as shown below:

\begin{equation}
\label{050}
\Gamma_{eff_{f-1-T}}(\omega t) = cos(\omega t) sin(\frac{\phi_{X} - \omega t}{2})cos(\frac{\phi_{X} + \omega t}{2})
\end{equation}

\subsection{Finding \(C_{0}\)}
\label{effISF-f-c0}
To find \(C_{0}\), which is the DC value of the effective ISF, the following formula stated in \cite{VCO-A1-7} is applied:

\begin{equation}
\label{051}
C_{0} = \frac{1}{\pi}\int_{0}^{2\pi} \Gamma_{eff} \, d{\omega t}
\end{equation}

Considering the specified work regions for the M1 transistor in figure \ref{fig-05}, the above integral equation can be decomposed into four parts as follows:

\begin{equation}
\label{052}
\begin{aligned}
C_{0_{M1}} &= \frac{1}{\pi} \left( \int_{0}^{\phi_{2}} \Gamma_{{eff}_{f-1-S}} \, d{\omega t} \right. \\
&\quad + \int_{\phi_{2}}^{\pi - \phi_{2}} \Gamma_{{eff}_{f-1-T}} \, d{\omega t} \\
&\quad + \int_{\pi - \phi_{2}}^{\pi + \phi_{1}} \Gamma_{{eff}_{f-1-S}} \, d{\omega t} \\
&\quad + \left. \int_{2\pi - \phi_{1}}^{2\pi} \Gamma_{{eff}_{f-1-S}} \, d{\omega t} \right)
\end{aligned}
\end{equation}

As a result of substituting Equations (\ref{046}) and (\ref{050}) into Equation (\ref{052}) and applying trigonometrical simplifications, the DC component of the effective ISF is updated as shown below.

\begin{equation}
\label{052+1}
\begin{aligned}
C_{0_{M1}} &= \frac{1}{2\pi} ( {sin}^2(\phi_{2}) - sin(\phi_{1}) sin(\phi_{X}) )
\end{aligned}
\end{equation}

According to ISF theory, a symmetric oscillation waveform helps suppress the up-conversion of flicker noise to the carrier frequency \cite{VCO-A1-18-1}. Therefore, assuming that the only-NMOS LC-VCO illustrated in Figure~\ref{fig-04} is symmetrically designed, the contributions of the flicker noise sources of M1 and M2 to the overall phase noise can be neglected. This implies that the DC value of the effective ISF, \(C_{0}\), is zero. Under this condition, the following expression can be inferred from Equation~(\ref{052+1}).

\begin{equation}
\label{052+2}
\begin{aligned}
sin(\phi_{X}) &= \frac{{sin}^2(\phi_{2})}{sin(\phi_{1})}
\end{aligned}
\end{equation}

Thus, if the condition given in Equation (\ref{052+2}) is met, flicker noise sources generated by the transistors will not contribute to the deterioration of the oscillator's overall phase noise.

\section{Phase Noise from Thermal Noise}
\label{ISF-th}
To determine how the thermal noise sources in an LC oscillator can affect phase noise, the following formula suggested by the ISF theory is applied:

\begin{equation}
\label{053}
L(\Delta\omega) \approx 10\log_{10}(\frac{\Gamma_{eff-rms}^2}{4q_{max}^2}\frac{\bar{i}_{n-th}^2}{\Delta f}\frac{1}{\Delta {\omega}^2})
\end{equation}

where \(\Gamma_{eff-rms}\) is the Root Mean Square (RMS) value of the effective ISF, \(q_{max}\) is the maximum charge across the LC tank, \(\bar{i}_{n-th}^2/ \Delta f\) is the power spectral density of thermal noise, and \(\Delta {\omega}\) represents the frequency offset for one source of thermal noise.
 
According to Equation (\ref{053}), minimizing the RMS value of the effective ISF of a CMOS device can help reduce phase noise caused by thermal noise. The first step in calculating phase noise caused by thermal noise in NMOS transistors is to determine their \(\Gamma_{eff-rms}\). Since the LC-VCO is assumed to be designed symmetrically, and the calculation results are identical for M1 and M2, only the \(\Gamma_{eff-rms}\) for M1 will be computed in order to reduce redundancy and complexity in the analysis.

\subsection{Calculating \(\Gamma_{eff-rms}\) due to Thermal Noise of M1}
\label{effISF-th}
Based on Equation (\ref{039}), to obtain the effective ISF caused by the thermal noise of the M1 transistor in figure \ref{fig-04}, the corresponding NMF affected by device thermal noise must first be determined. According to \cite{VCO-A1-24}, the following formula can be used to derive the thermal noise modulating function of a CMOS transistor:

\begin{equation}
\label{054}
\alpha_{th_{1}}(\omega t) = \sqrt{\frac{\bar{i}_{n-th}^2}{max(\bar{i}_{n-th}^2)}}
\end{equation}

where \(\alpha_{th_{1}}(\omega t)\) is the thermal-noise modulating function, and \({\bar{i}_{n-th}^2}/\Delta f\) is the thermal-noise power spectral density in M1. As a result of substituting Equation (\ref{016}) into Equation (\ref{054}), the NMF of M1 resulting from thermal noise in saturation mode is updated as follows:

\begin{equation}
\label{055}
\alpha_{th_{1-S}}(\omega t) = \sqrt{\frac{V_{GS1} + nV_{BS1} - V_{Th0}}{max(V_{GS1} + nV_{BS1} - V_{Th0})}}
\end{equation}

In saturation mode, by applying the assumptions stated in Equation~(\ref{019}) to Equation~(\ref{055}) and performing mathematical simplification, the normalized thermal noise NMF of M1 is obtained as:

\begin{equation}
\label{056}
\alpha_{th_{1-S}}(\omega t) = \sqrt{sin(\frac{\phi_{1} - \omega t}{2})cos(\frac{\phi_{1} + \omega t}{2})}
\end{equation}

When Equations~(\ref{056}) and (\ref{040}) are substituted into Equation~(\ref{039}), the effective ISF due to the thermal noise of M1 in saturation mode is obtained as:

\begin{equation}
\label{057}
\Gamma_{eff_{th-1-S}} = cos(\omega t) \sqrt{sin(\frac{\phi_{1} - \omega t}{2})cos(\frac{\phi_{1} + \omega t}{2})}
\end{equation}

In the cut-off region of M1, the device transconductance is nearly zero; therefore, the PSD of thermal noise in this mode can be approximated as negligible. As a result, computing the effective ISF in this region is unnecessary. In contrast, in the triode region, the NMF of M1 can be obtained by substituting Equation~(\ref{018}) into Equation~(\ref{054}), as expressed below:

\begin{equation}
\label{058}
\alpha_{th_{1-T}}(\omega t) = \sqrt{\frac{V_{GS1} + n(V_{DS1} + V_{BS1}) - V_{Th0}}{max(V_{GS1} + n(V_{DS1} + V_{BS1}) - V_{Th0})}}
\end{equation}

Assuming that, in an only-NMOS cross-coupled LC-VCO, both the AC amplitude and the DC level of the output oscillating signals in steady state are approximately equal to \(V_{DD}\), and by applying the assumptions in Equation~(\ref{019}) to Equation~(\ref{058}), followed by trigonometric simplifications, the normalized thermal noise NMF for M1 in triode mode is given as:

\begin{equation}
\label{059}
\alpha_{th_{1-T}}(\omega t) = \sqrt{sin(\frac{\phi_{X} - \omega t}{2})cos(\frac{\phi_{X} + \omega t}{2})}
\end{equation}

where \(\phi_{X}\) is the same angle as defined by Equation (\ref{049}). As a result of substituting Equations (\ref{059}) and (\ref{040}) into (\ref{039}), the effective ISF due to thermal noise in the triode mode of M1 is calculated as follows:

\begin{equation}
\label{060}
\Gamma_{eff_{th-1-T}} = cos(\omega t)\sqrt{sin(\frac{\phi_{X} - \omega t}{2})cos(\frac{\phi_{X} + \omega t}{2})}
\end{equation}

\subsection{Calculating the RMS value of the \(\Gamma_{eff}(\omega t)\)}
\label{effISF-th-rms}
According to \cite{VCO-A1-7}, to find the RMS value of \(\Gamma_{eff}\), the following formulas can be used:

\begin{equation}
\label{061}
{\Gamma^2_{eff-RMS}} = \frac{1}{2\pi}\int_{0}^{2\pi}\left| \Gamma_{eff}(\omega t)\right|^2 \, d{\omega t}
\end{equation}

Based on the operating regions of M1 over a \(2\pi\) oscillation period, as illustrated in Figure~\ref{fig-05}, the integral above can be partitioned into four segments, as shown below:

\begin{equation}
\label{062}
\begin{aligned}
{\Gamma^2_{eff-RMS_{1}}}  &= \frac{1}{2\pi} \left( \int_{0}^{\phi_{2}} \left|\Gamma_{{eff}_{th-1-S}}\right|^2 \, d{\omega t} \right. \\
&\quad + \int_{\phi_{2}}^{\pi -\phi_{2}}\left|\Gamma_{{eff}_{th-1-T}}\right|^2 \, d{\omega t} \\
&\quad + \int_{\pi - \phi_{2}}^{\pi + \phi_{1}} \left|\Gamma_{{eff}_{th-1-S}}\right|^2 \, d{\omega t} \\
&\quad + \left. \int_{2\pi - \phi_{1}}^{2\pi} \left|\Gamma_{{eff}_{th-1-S}}\right|^2 \, d{\omega t} \right)
\end{aligned}
\end{equation}

By substituting \(\Gamma_{\text{eff}_{\text{th-1-S}}}\) from Equation~(\ref{057}) and \(\Gamma_{\text{eff}_{\text{th-1-T}}}\) from Equation~(\ref{060}), and performing mathematical simplifications, the following expression is obtained for the squared RMS value of the effective ISF of the M1 transistor:

\begin{equation}
\label{063}
{\Gamma^2_{eff-RMS_{1}}} = \frac{1}{8\pi}\left( sin\phi_{1} f(\phi_{1},\phi_{2}) +  sin\phi_{X} g(\phi_{2}) \right)
\end{equation}

where \(f(\phi_{1},\phi_{2})\) is 

\begin{equation}
\label{064}
f(\phi_{1},\phi_{2}) = 2\phi_{1} + 2\phi_{2} + \frac{5}{3}sin2\phi_{1} + sin2\phi_{2} - \frac{4}{3}cot\phi_{1}
\end{equation}

and \(g(\phi_{2})\) is

\begin{equation}
\label{065}
g(\phi_{2}) = \pi - 2\phi_{2}  - sin2\phi_{2}
\end{equation}

\section{Phase Noise Minimization}
\label{PN-min}

As discussed in Section~\ref{ISF-f}, the flicker noise sources of M1 and M2 do not contribute to the output phase noise when the DC component of the effective ISF, \(C_0\), is zero. However, according to Section~\ref{ISF-th}, the thermal noise of the transistors significantly impacts phase noise performance. As indicated in Equation~(\ref{063}), the RMS value of the effective ISF is a function of three variables: \(\phi_1\), \(\phi_2\), and \(\phi_X\). In the LC-VCO circuit shown in Figure~\ref{fig-04}, variations in circuit parameters, such as changes to the amplitude and DC level of the output oscillation, or the body bias signal applied to the transistors, can influence these phase angles. Therefore, tuning \(\phi_1\), \(\phi_2\), and \(\phi_X\) is advantageous for minimizing the RMS value of the effective ISF. Based on Equations~(\ref{064}) and~(\ref{065}), setting \(\phi_2 = 90^\circ\) effectively eliminates the triode operation regions of M1 and M2. As a result, the function \(g(\phi_2)\) becomes zero, simplifying Equation~(\ref{063}) as shown below:

\begin{equation}
\label{066}
{\Gamma_{eff-RMS_{1}}}^2 = \frac{1}{8\pi}sin\phi_{1} f(\phi_{1},\phi_{2})
\end{equation}

where function \(f(\phi_{1},\phi_{2})\) is also modified as follows:

\begin{equation}
\label{067}
f(\phi_{1},\phi_{2}) = \pi + 2\phi_{1} + \frac{5}{3}sin2\phi_{1} - \frac{4}{3}cot\phi_{1}
\end{equation}

According to Equation (\ref{066}), there are two theoretical solutions to eliminate the thermal noise effects of the cross-coupled transistors on the phase noise performance. Firstly, when \(\phi_{1}\) is set to zero, the RMS value of the effective ISF becomes zero. In practice, however, it is not feasible, since saturation regions would also be eliminated under this condition. This would cause the oscillator to stop oscillating due to insufficient negative resistance seen by the LC tank. The second method is to set the function \(f(\phi_{1},\phi_{2})\) equal to zero resulting in a zero RMS value for the effective ISF. As a result, equation (\ref{067}) is set to zero as follows:

\begin{equation}
\label{068}
\pi + 2\phi_{1} + \frac{5}{3}sin2\phi_{1} - \frac{4}{3}cot\phi_{1} = 0
\end{equation}

By employing numerical methods to solve the Equation (\ref{068}), the following non-zero answer for \(\phi_{1}\) in M1 transistor over an oscillation period of \(2\pi\) is approximated:

\begin{equation}
\label{069}
\phi_{1} \approx 16.172^\circ
\end{equation}

Thus, to minimize the thermal noise impact of the cross-coupled transistors, \(\phi_{1}\) and \(\phi_{2}\) must be set at 16.172° and 90°, respectively. The following changes are made to Equations (\ref{027}) and (\ref{028}) as a result of taking these two angle values into account:

\begin{equation}
\label{070}
16.172^\circ = sin^{-1}(\frac{V_{Th0} - nV_{DC1} - V_{DC0}}{A - nA_{1}})
\end{equation}

\begin{equation}
\label{071}
90^\circ = sin^{-1}(\frac{V_{Th0} - nV_{DC1}}{2A - nA_{1}})
\end{equation}

By combining these two equations and performing algebraic simplifications, the AC amplitude and DC offset of the symmetric signals applied to the body terminals of the cross-coupled transistors can be expressed as follows:

\begin{equation}
\label{072}
A_{1} = \frac{1}{n}\left( A +  \frac{A - V_{DC0}}{1 - sin(16.172^\circ)} \right)
\end{equation}

\begin{equation}
\label{073}
V_{DC1} = \frac{1}{n}\left( V_{Th0} +  \frac{A.sin(16.172^\circ) - V_{DC0}}{1 - sin(16.172^\circ)} \right)
\end{equation}

By substituting Equation (\ref{04}) into Equations (\ref{072}) and (\ref{073}), the resulting expressions can be reformulated as follows:

\begin{equation}
\label{074}
A_{1} = \frac{\sqrt {\Phi_{s}}}{\gamma}\left( A +  \frac{A - V_{DC0}}{1 - sin(16.172^\circ)} \right)
\end{equation}

\begin{equation}
\label{075}
V_{DC1} = \frac{\sqrt {\Phi_{s}}}{\gamma}\left( V_{Th0} +  \frac{A.sin(16.172^\circ) - V_{DC0}}{1 - sin(16.172^\circ)} \right)
\end{equation}

Equations (\ref{074}) and (\ref{075}) can be used to observe how the circuit parameters in the body-biased only-NMOS LC-VCO shown in Figure \ref{fig-04} interact with each other to minimize phase noise caused by cross-coupled devices. In order to justify these equations, the circuit parameters should be tuned during the design process in order to avoid deterioration of the output phase noise caused by the internal noise of the cross-coupled transistors.

\section{Closed-Form Expressions}
In the last section, two closed-form expressions have been derived for the amplitude and the DC components of the signals that are appled to the body terminals of the crossed-coupled NMOS transistors. Now, in this section, the application of the Equations (\ref{074}) and (\ref{075}) is examined. By assuming that \(A\) and \(V_{DC0}\) in an only NMOS LC-VCO are approximately equal to the value of the DC power supply, \(V_{DD}\), the expressions can be simplified as follows:

\begin{equation}
\label{076}
A_{1} \approx \frac{\sqrt {\Phi_{s}}}{\gamma} V_{DD} 
\end{equation}

\begin{equation}
\label{077}
V_{DC1} \approx \frac{\sqrt {\Phi_{s}}}{\gamma}\left( V_{Th0} -  V_{DD} \right)
\end{equation}

Dividing the Equation (\ref{077}) by Equation (\ref{076}) results in the following expression.

\begin{equation}
\label{078}
\left| \frac{V_{DC1}}{A_{1}} \right| = \left| \frac{V_{Th0}}{V_{DD}} - 1 \right|
\end{equation}

Equation (\ref{078}) demonstrates how an only-NMOS oscillator should be designed to minimize phase noise due to thermal noise and flicker noise of the cross-coupled transistors.

\section{Modeling and Simulation}

\begin{figure}[]
\centering
\noindent
\includegraphics[width=3.5in, trim=0cm 0cm 0cm 0cm, clip]{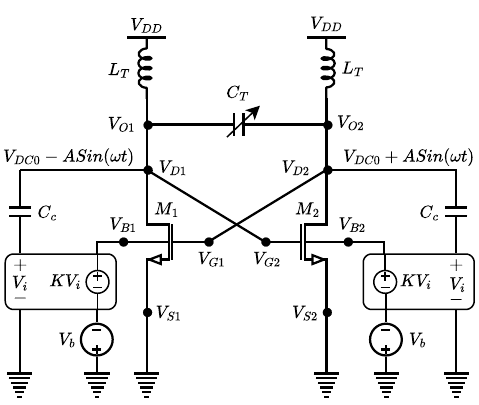}
\caption{Simulink model for the realization of an ultra-low phase noise, body-biased, NMOS-only cross-coupled LC-VCO.}
\label{fig-06}
\end{figure}

One possible implementation of Equation~(\ref{078}) involves establishing a feedback path from the oscillator outputs to the body terminals of the cross-coupled transistors. In this configuration, it is assumed that an attenuation stage is utilized to reduce the AC amplitude of the output oscillating signals before they are applied to the bulk terminals of the transistors. As illustrated in Figure~\ref{fig-06}, the DC component of the output signals is approximately blocked by the coupling capacitors \(C_{c}\). Subsequently, a DC bias is added to the attenuated signals prior to their application to the body terminals, enabling control over the DC level.

The final signals applied to the body terminals of transistors M1 and M2 can be mathematically expressed as follows:

\begin{equation}
\label{079}
V_{DC1} \approx -V_{b}
\end{equation}

\begin{equation}
\label{080}
A_{1} = KA  \approx K V_{DD}
\end{equation}

Here, $K$ denotes the attenuation factor, and $V_{b}$ represents the added DC bias used to shift the DC component of the body signals. This adjustment is crucial for optimizing the body bias when phase noise minimization is targeted. By substituting Equations~(\ref{079}) and~(\ref{080}) into Equation~(\ref{078}), a refined closed-form expression can be derived that reveals the interdependence among the threshold voltage ($V_{th0}$), power supply voltage ($V_{DD}$), attenuation factor ($K$), and the tuning DC bias voltage ($V_{b}$). 

\begin{equation}
\label{081}
\left| \frac{V_{b}}{KV_{DD}}\right| = \left| \frac{V_{Th0}}{V_{DD}} - 1 \right|
\end{equation}

\begin{figure}[]
\centering
\noindent
\includegraphics[width=\linewidth, trim=0cm 0cm 0cm 0cm, clip]{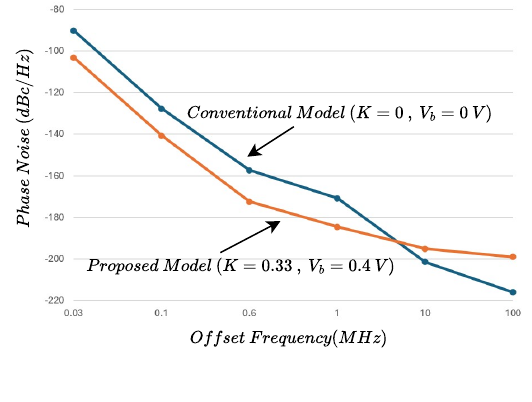}
\caption{Simulated phase noise comparison between the conventional LC-VCO and the proposed body-biased NMOS-only LC-VCO model across various offset frequencies.}
\label{fig-07}
\end{figure}

To simulate Equation~(\ref{081}), the model illustrated in Figure~\ref{fig-06} is implemented using MATLAB and Simulink. A supply voltage of 1.8~V, commonly used in integrated circuit (IC) design, is selected as a practical value for \(V_{DD}\). Based on standard CMOS process parameters, a threshold voltage of \(V_{\text{th0}} = 0.5~\text{V}\) is assumed. An attenuation factor of \(K = 0.33\) and a DC body bias voltage of \(V_{b} = 0.4~\text{V}\) are adopted, as they lead to enhanced phase noise performance according to simulation outcomes. Phase noise is measured at several frequency offsets from the carrier, specifically at 10~kHz, 30~kHz, 600~kHz, 1~MHz, 10~MHz, and 100~MHz. Figure~\ref{fig-07} presents the phase noise performance of both the conventional LC-VCO (with \(K = 0\) and \(V_{b} = 0\)) and the proposed LC-VCO (with \(K = 0.33\) and \(V_{b} = 0.4~\text{V}\)) across these offsets. The results indicate a significant phase noise improvement for frequency offsets up to approximately 10~MHz. Specifically, the proposed design achieves an improvement of approximately 12~dB at lower offset frequencies and nearly 15~dB around a 1~MHz offset, demonstrating a considerable reduction in phase noise due to the suppression of both flicker and thermal noise components.

\section{Conclusion}
\label{Con}

In this article, a comprehensive phase noise analysis based on the ISF theory was presented using a three-terminal transistor model. By analyzing the flicker and thermal noise contributions from active devices, two useful closed-form expressions for the amplitude and DC offset of the first harmonic of the body signals were derived. These formulas enable the design of various circuit topologies for an ultra-low phase noise, NMOS-only, dynamically body-biased, cross-coupled LC-VCO. Satisfying the conditions dictated by these formulas can effectively ensure that noise generated by the active devices has minimal impact on the overall phase noise. Under such conditions, the dominant factor affecting phase noise becomes the quality factor of the LC tank. This work provides valuable insights for RFIC circuit designers seeking to minimize phase noise in oscillator architectures. Moreover, incorporating more accurate models in future simulations and analyses may lead to further improvements, particularly in phase noise performance at higher offset frequencies.

\appendix

\section{Work Regions of M2}
\label{M2-Regions}

\subsection{Saturation Region of M2}
In the case of the M2 transistor, since the drain, gate, and body pins are controlled by \(V_{O1}\), \(V_{O2}\), and \(V_{B2}\), the saturation criteria can be updated as follows:

\begin{equation}
\label{116}
V_{O1} \geq V_{Th0} - nV_{B2}
\end{equation}

\begin{equation}
\label{117}
V_{O2} + (V_{Th0} - nV_{B2}) \geq V_{O1}
\end{equation}

By applying Equation (\ref{020}) to Equations (\ref{116}) and (\ref{117}) as well as some simplifications, the following criteria must be satisfied for M2 to operate at its saturation state.

\begin{equation}
\label{118}
sin(\omega t) \leq \frac{V_{DC0} - V_{Th0} + nV_{DC1}}{A - nA_{1}} 
\end{equation}

\begin{equation}
\label{119}
sin(\omega t) \geq \frac{nV_{DC1} - V_{Th0}}{2A - nA_{1}} 
\end{equation}

Equations (\ref{027}) and (\ref{028}) are substituted into Equations (\ref{118}) and (\ref{119}), resulting in the following equations:

\begin{equation}
\label{120}
sin(\omega t) \leq sin(-\phi_{1})
\end{equation}

\begin{equation}
\label{121}
sin(\omega t) \geq sin(-\phi_{2})
\end{equation}

As a result of combining equations (\ref{120}) and (\ref{121}), the following expression for the saturation mode of the M2 transistor can be obtained.

\begin{equation}
\label{122}
sin(-\phi_{2}) \leq sin(\omega t) \leq sin(-\phi_{1})
\end{equation}

After solving Equation (\ref{122}) over a \(2\pi\) period, the following ranges are obtained for M2 in saturation mode.

\begin{equation}
\label{123}
\left\{
\begin{array}{ll}
0 \leq \omega t \leq \phi_{1} \\
\pi - \phi_{1} \leq \omega t \leq \pi + \phi_{2} \\
2\pi - \phi_{2} \leq \omega t \leq 2\pi
\end{array}
\right.
\end{equation}

\subsection{Triode Region of M2}
\label{M2-T}
Triode mode requires the M2 to be turned on, and its drain-source voltage must be lower than its overdrive voltage (\(V_{GS} - V_{Th}\)). Consequently, the following conditions can be derived from updating equations (\ref{118}) and (\ref{119}) for this region:

\begin{equation}
\label{124}
sin(\omega t) \leq \frac{V_{DC0} - V_{Th0} + nV_{DC1}}{A - nA_{1}} 
\end{equation}

\begin{equation}
\label{125}
sin(\omega t) < \frac{nV_{DC1} - V_{Th0}}{2A - nA_{1}} 
\end{equation}

By substituting Equations (\ref{027}) and (\ref{028}) into Equations (\ref{124}) and (\ref{125}), the above equations are updated as follows:

\begin{equation}
\label{126-1}
sin(\omega t) \leq sin(-\phi_{1})
\end{equation}

\begin{equation}
\label{126-2}
sin(\omega t) < sin(-\phi_{2})
\end{equation}

As a result of solving equations (\ref{126-1}) and (\ref{126-2}), the following range represents the region over which M2 operates in triode mode.

\begin{equation}
\label{127}
\pi + \phi_{2} < \omega t < 2\pi - \phi_{2}
\end{equation}

\subsection{Cut-Off Region of M2}
\label{M2-T}
In the cut-off mode, M2 has a gate-source voltage below its threshold voltage. Consequently, equation (\ref{118}) is updated for this region in the following manner:

\begin{equation}
\label{128}
sin(\omega t) > \frac{V_{DC0} - V_{Th0} + nV_{DC1}}{A - nA_{1}} 
\end{equation}

Taking equation (\ref{027}) into account, equation (\ref{128}) is revised as follows:

\begin{equation}
\label{129}
sin(\omega t) > sin(-\phi_{1})
\end{equation}

Solving equation (\ref{129}), the following range identifies the area within which M2 operates in cut-off mode.

\begin{equation}
\label{130}
\phi_{1} < \omega t < \pi - \phi_{1}
\end{equation}

\bibliographystyle{elsarticle-num} 
\bibliography{MyBibDatabase}






\end{document}